\documentclass[prb,aps,twocolumn,floats,epsfig,showpacs]{revtex4}
\usepackage {graphicx}
\begin{document}
\draft
\title{\bf Model Hessian for accelerating first-principles
structure optimizations }
\author{M. V Fern\'andez-Serra,$^{1,2}$ \ Emilio Artacho$^{1}$,
and Jos\'e M. Soler$^2$}
\address{
$^1$Department of Earth Sciences, University of Cambridge, Downing
Street, Cambridge CB2 3EQ, UK\\
$^2$Depto. de F\'{\i}sica de la Materia Condensada, C-III,
Universidad Aut\'onoma, 28049 Madrid, Spain}
\date{\today}

\begin{abstract}

We present two methods to accelerate first-principles structural
relaxations, both based on the dynamical matrix obtained from a
universal model of springs for bond stretching and bending.
Despite its simplicity, the normal modes of this model Hessian
represent excellent internal coordinates for molecules and solids
irrespective of coordination, capturing not only the
long-wavelength acoustic modes of large systems, but also the
short-wavelength low-frequency modes that appear in complex
systems. In the first method, the model Hessian is used to
precondition a conjugate gradients minimization, thereby
drastically reducing the effective spectral width and thus
obtaining a substantial improvement of convergence. The same
Hessian is used in the second method as a starting point of a
quasi-Newton algorithm (Broyden's method and modifications
thereof), reducing the number of steps needed to find the correct
Hessian. Results for both methods are presented for geometry
optimizations of clusters, slabs, and biomolecules, with 
speed-up factors between 2 and 8.
\end{abstract}



\maketitle


  The exponential growth of computer power together with
recent methodological advances have opened first-principles
electronic-structure calculations to systems of unprecedented size
and complexity.
  Some first-principles linear-scaling methods\cite{Jose,Gillan}
(methods for which the required computer resources scale linearly
with system size) already allow calculations involving a few
thousand atoms.
  In many cases, the first problem to solve when facing a large and
complex system is the first-principles determination of its
structure by means of some energy minimization algorithm.
  As pointed out by Goedecker {\it et al.}\cite{Goedecker}, the
linear scaling achieved for each ab initio force evaluation 
degrades to a higher order scaling for the structure determination,
since the number of force evaluations increases with system size.
  This is due to the increasing ill-conditioning of the structural
optimization with larger system sizes, which has become the main
bottleneck in the ab initio prediction of structures for the sizes
treatable now.

  In the case of conjugate gradients, for example, the number of
evaluations is proportional to the condition number, defined as
the ratio between the highest and the lowest non-zero curvatures
of the energy landscape around the sought minimum.
  In any solid system, in the limit of large sizes, the lowest
non-zero frequency is inversely proportional to a characteristic
side length of the system, given the linear dispersion relation of
long wave-length acoustic phonons.
  The condition number and thus the number of relaxation steps
contributes an additional overall scaling factor of between 
$N^{1/3}$ and $N$, depending on the system shape.
  A clever solution \cite{Goedecker} is to use the known (or
approximate) macroscopic elastic properties to relax the 
long-wavelength low-frequency modes.
  However, with an increasing number of atoms, there is also an
increase in the complexity of the system and of the different kinds 
of low frequency modes.
  In many complex systems of enormous importance
(polymers, biomolecules, glasses, to name a few) there are
low-frequency modes that do not correspond to any long-wavelength
continuum limit.

  This ``complexity" is best characterized within the theory of
rigidity\cite{rigid} and its floppy modes.
  Their frequencies are very low since they do not involve the
stretching, bending, or torsion of any particular bond, but their 
wavelengths can be very short, in many instances the modes being 
localized.
  A system with floppy modes is therefore ill-conditioned for
relaxation.
  Even if these floppy modes do not strictly enter scaling
arguments, they do spoil the structural optimizations of complex
systems.
  The language introduced naturally in rigidity theory is that
of internal coordinates: bond lengths, bond angles, and bond
torsion angles.
  Internal coordinates are quite popular in chemistry and have
indeed proven to be more efficient than Cartesian coordinates for
the optimization of molecular systems.\cite{Pulay,Baker}
  They are weaker, however, for large condensed systems since, on
one hand, they do not address the acoustic ill-conditioning and,
on the other hand, they are more complicated to handle 
for high coordinations \cite{Andzelm}.

  It is important to stress that avoiding ill conditioning
requires a non-pathological identification of the low 
frequency modes, more than a very realistic description of
the dynamical matrix.
  Realistic empirical potentials, no matter how good or universal
\cite{uff}, provide good dynamical matrices close to the minimum
but may yield negative curvatures away from it.
  We present here a simple and natural way of addressing both
sources of ill conditioning (acoustic and floppy) on the same
footing.
  It constructs a positive-definite dynamical matrix from a universal 
model of springs for bond stretching and bending, and uses it to 
improve the geometry relaxation in two alternative ways, which 
correspond to two popular minimization methods.


{\em Model Potential.}
  The proposed Hessian model is based on a simple bond bending and 
stretching potential defined for any system at the given coordinates 
${\bf r}^0_i$ as
\begin{equation}
U=\frac{1}{2}\sum_{i<j} k^s_{ij}(r_{ij}-r_{ij}^{\rm 0})^2\,
+\frac{1}{2}\sum_{i<j<l}k^b_{ijl}(\theta_{ijl}-
\theta_{ijl}^{\rm 0})^2,
\end{equation}
\begin{equation}
k^s_{ij} = A \left ( \frac{R_i+R_j}{r_{ij}^{\rm 0} } \right )^8,
\end{equation}
\begin{equation}
k^b_{ijk} = B \sqrt{k^s_{ij} k^s_{jk}} r^0_{ij} r^0_{jk} 
\end{equation}
where $R_i$ and $R_j$ are the covalent radii of the atoms
connected\cite{covalent}. The sums in Eq.~(1) are limited to neighbors within
$6\times max_{i}[R_i]$
  The proportionality constant $A$ in Eq.~(2) is irrelevant for the
the first relaxation method described below but it does affect
the second one (very moderately).
  Its value $A=3.0 \times 10^5 $ eV/{\AA$^2$} has been defined universally by adjusting 
the bulk modulus of cubic diamond, the largest known in nature.
  Albeit arbitrary, a large constant ensures small initial atomic
displacements and thus the stability of the relaxation.
  The power 8 is also arbitrary but reasonable and has been 
chosen after some numerical tests.
  The relative magnitude of the bending and stretching constants 
has also been arbitrarily chosen as $B=1/10$ after some tests.
  The introduction of the even weaker torsional forces in the 
potential could bring further benefits and will be explored in 
later works.

   The Hessian is evaluated at the minimum of the potential, which 
is defined by Eq.~(1) to be {\em at the given (initial) coordinates} 
of the system to be relaxed.
   A positive definite Hessian is thus guaranteed.
   The model has no system-dependent parameters and is completely universal.
   Despite its simplicity, it captures qualitatively the separation
between high-frequency stretching modes and lower-frequency 
bending and torsion modes.
   It also yields naturally the long wavelength acoustic modes, thus
combining the main advantages of the methods of internal
coordinates \cite{Pulay,Baker} and of elastic modes \cite{Goedecker}.

{\em Preconditioned conjugate gradients.}
  The problem of ill conditioned minimization can be seen as 
the difficulty to find the way to the minimum along a gently 
sloped but narrowly shaped valley.
  The problem can be quantified by the condition number, 
the ratio $\omega^2_{max}/\omega^2_{min}$ between the curvatures 
across and along the valley or, in a higher dimensional space, 
between the largest and smallest curvatures.
  The number of steps in the minimization process increases with
growing condition number, the precise scaling depending on the
particular algorithm, quadratic for steepest descent and linear
for conjugate gradients, for example.

  The number of evaluations of the ab initio forces (the gradient)
can be thus reduced by a transformation of coordinates such that
the curvatures in the new space give a smaller condition number.
  The effort required for the coordinate transformations is 
negligible compared with that of calculating the ab initio forces.
  Such preconditioning can be accomplished by using a priori knowledge
about the system curvatures.

  Our preconditioning procedure requires an
initial diagonalization of the Hessian matrix of the model defined
in Eq. (1):
\begin{equation}
H_{ij}=\frac{\partial^2U}{\partial x_i\partial x_j} .
\end{equation}
  The ab initio forces ${\bf F}$ are then projected over the normal
modes ${\bf n}$ of that Hessian matrix and divided by the `frequencies' 
$\omega_n$ (the square root of the eigenvalues):
\begin{equation}
f_n = \frac{{\bf F} \cdot {\bf n}}{\omega_n}
\end{equation}
  A conventional conjugate gradients routine is then used to minimize
the ab initio energy as a function of the new coordinates:
\begin{equation}
y_n = \omega_n ({\bf x} \cdot {\bf n})
\end{equation}
where {\bf x} are the Cartesian coordinates.

  The initial diagonalization represents an unimportant
computational effort for the systems treated nowadays with
first-principles techniques.
  In the future, however, this foreseeable limiting step will have
to be given some more thought.
  For example, the mentioned rigidity theory could be further
exploited, given its ability of predicting subsets of floppy atoms
versus rigid bits of the overall structure.\cite{rigid}

\begin{figure}
\includegraphics{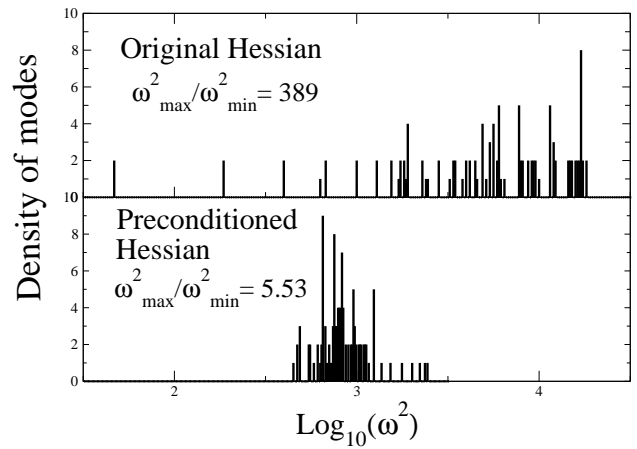}
\caption[1]{\label{Spectral width}
Histogram of eigenvalues $\omega^2$ of the Hessian (matrix of 
second derivatives) of a 247-atom gold slab made of
ten layers of 25 atoms each, with three vacancies.
simulated with an embedded atom potential \cite{embeded}. 
Upper panel: using standard Cartesian coordinates. 
Lower panel: using transformed coordinates, for which the 
eigenvalues of our model Hessian become all equal.
The condition number (ratio between largest and smallest 
eigenvalues) is given for each case.
The units of $\omega$ are cm$^{-1}$.}
\end{figure}

  Fig.~\ref{Spectral width} shows the reduction in spectral width 
and condition number for a 10-layer gold slab,
simulated with an embedded atom potential \cite{embeded}.
  This many-body potential has been chosen for some of our tests
because the efficiency of our minimization methods, in terms of 
the number of iterations, does not depend on the specific form of 
the interactions.
  In the new coordinates
the condition number is reduced by a factor 70, proving that the model 
potential represents a good initial approximation to the real one.

{\em Quasi-Newton method.}
  Variable-metric methods\cite{recipes,Vanderbilt} minimize the energy 
by applying
\begin{equation}
\delta  {\bf x} = - H^{-1} \delta {\bf F} = H^{-1} {\bf F}
\label{Broyden}
\end{equation}
where $H$ is the Hessian and $\delta{\bf F} = -{\bf F}$ is the
desired change of the forces.
  Given the exact Hessian of a perfectly harmonic function, one step
would suffice to find its minimum.
  Since the potential is generally not harmonic and the Hessian is 
unknown, an iterative process is followed in practice, starting with 
a trial Hessian, moving according to (\ref{Broyden}) and
updating the Hessian afterwards so that it obeys (\ref{Broyden}) 
for the true change of the force found.
  Since no information is generally known about the Hessian,
a unity matrix is used as the initial $H^{-1}$.
  We simply change this to the inverse of our model Hessian
and then use a standard variable-metric method \cite{Vanderbilt}.
  Again, the fact that the model Hessian captures non-trivial
low-frequency modes allows the iterative learning process to save
many steps to find out about them (the ill-conditioning affects 
this kind of methods as well).
  The initial inversion of the model Hessian represents a very
small overhead compared with the evaluation of ab initio forces.

  In addition, Broyden-like methods lend themselves nicely to
hierarchical approaches whereby lower quality relaxations can feed
the Hessian for higher quality ones.
  For example, one can perform a relaxation with a small basis set
for the electronic structure, starting from our model Hessian, and 
use the resulting ab initio inverse Hessian to launch a
finer relaxation with a better basis set.
  We will explore this approach in a later work.


  The two methods have been applied to systems of different
character and complexity, namely, to metal clusters and slabs, and
a biomolecule.
  They constitute quite extreme examples of complex systems.
  On one hand, two systems with high coordinations and a non trivial
energy landscape.\cite{prb-au-clus}
  On the other, a low-coordination molecule with floppy modes.

  Table I shows results for the relaxation of two gold systems, 
simulated with the embedded atom potential~\cite{embeded}: 
a ten-layer crystalline slab and an amorphous 77-atom cluster.
  The number of iterations needed to reach the minimum are 
given as a function of its distance from the initial coordinates, 
chosen at random within a window of $\pm \delta x^0$ from the 
relaxed structure.
  Our two methods are compared
with (i) regular CG, (ii) CG preconditioned with the exact Hessian, 
and (iii) Broyden's method starting from the identity matrix.
The exact Hessian is calculated at the minimum because otherwise
it soon develops negative eigenvalues, whereas the model
one is evaluated at the starting point, since it is always
positive definite.

  The table shows that our model Hessian improves considerably
the efficiency of both the CG and the modified Broyden methods.
  As expected, the efficiency of the Broyden-like schemes is
superior when starting close to the minimum, though this advantage
decreases with initial separation from the harmonic basin.
  It is also interesting the comparison of our preconditioned CG
with the one using the exact Hessian at the minimum
(which is generally not available in practice, of course).
  The latter is extraordinarily efficient when started well within 
the harmonic basin, but it deteriorates very rapidly with distance, 
making it not much better than our method in practice.
  We do not, therefore, expect a better performance if using
realistic empirical potentials instead of our universal model.

\begin{table*}
\caption[ ]{
  Number of iterations needed to reach a force tolerance of 
10$^{-6}$ eV/{\AA} in a ten-layer crystalline slab and a 77-atom 
cluster of gold, simulated with an embedded atom potential.
  The initial atomic coordinates were randomly displaced from 
the equilibrium geometry in an interval of $\pm \delta x^0$.
  $F^0_{max}$ is the maximum initial atomic force.
  The conjugate gradient method was used with conventional
Cartesian coordinates (CG) and with preconditioned coordinates 
(PGC) that diagonalize the model Hessian (at the initial geometry) 
or the exact Hessian (at the minimum).
  The variable-metric Broyden method was used starting with the
conventional unit-matrix Hessian and with our new model Hessian.
}
\begin{ruledtabular}
\begin{tabular}{cccccccc}
System & $\delta x^0$ & $F^0_{max}$ & CG & PCG & PCG & Broyden  &
Broyden  \\
  & (\AA) & (eV/\AA) & & Model H & Exact H & Unit H  & Model H \\
\hline
 250-atom   & 10$^{-4}$ & $2.13 \times 10^{-3}$ &  40 & 18 &  3 &  8 &  5 \\
crystalline & 10$^{-3}$ & $2.13 \times 10^{-2}$ & 100 & 29 &  7 & 15 &  9 \\
   slab     & 10$^{-2}$ & $2.14 \times 10^{-1}$ & 180 & 35 & 10 & 32 & 14 \\
            & 10$^{-1}$ &  2.52                 & 240 & 60 & 45 &339 & 33 \\
\hline
 77-atom    & 10$^{-4}$ & $3.03 \times 10^{-3}$ &  51 & 22 &  3 & 22 & 14 \\
amorphous   & 10$^{-3}$ & $3.03 \times 10^{-2}$ &  72 & 40 &  6 & 29 & 19 \\
 cluster    & 10$^{-2}$ & $3.04 \times 10^{-1}$ & 101 & 53 &  9 & 36 & 25 \\
            & 10$^{-1}$ &  3.25                 & 167 & 90 & 50 &140 & 27 \\
\end{tabular}
\end{ruledtabular}
\label{77-cluster}
\end{table*}

\begin{figure}
\includegraphics[width=\columnwidth]{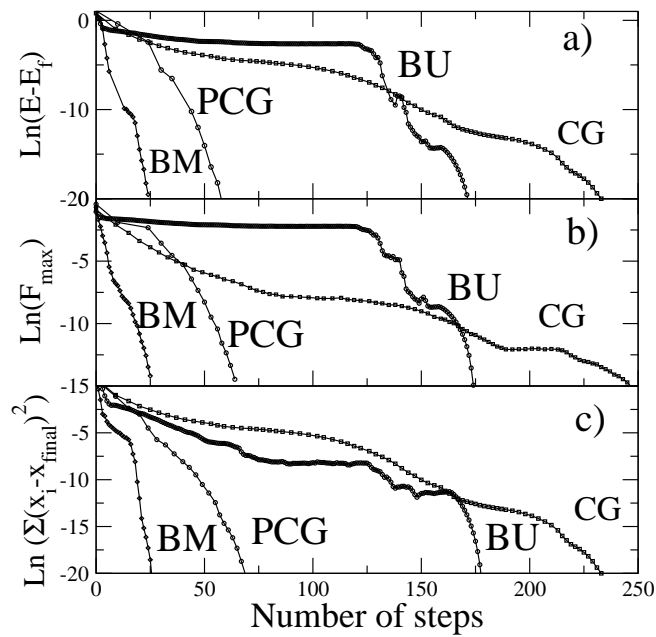}
\caption[2]{\label{convergence} 
   Logarithmic convergence of the energy (a), forces (b), and 
coordinates (c), for a 10-layer gold slab with 247 atoms and
three vacancies. 
   The initial coordinates where randomly displaced
by $\sim 0.2$~{\AA}  from the equilibrium geometry.
   Four different minimization methods were used:
conventional conjugate gradients (CG);
preconditioned conjugate gradients (PGC);
Broyden's method starting from a unit Hessian matrix (BU); and
Broyden's method starting from our model Hessian (BM).
} 
\end{figure}

  Fig.~\ref{convergence} shows the convergence in energy, forces, and 
atomic positions, for the ten-layer gold slab with the different 
methods discussed.
  Within CG, the convergence is slower during the first steps with the 
preconditioned method for the energy and the forces, but not for the 
coordinates.
  This is because CG responds with larger displacements to the high
curvature modes, which dominate the energy drop in the initial stages.
  The situation is reverted soon, however, and the overall efficiency
is clearly better for preconditioned CG.

  The modified Broyden method with our model Hessian initialization
is most efficient for small displacements from the minimum,
accelerating the convergence by factors of 3 to 8.
  In this and other systems, we have generally found that the Broyden 
method is extremely effective whenever it starts well within the harmonic 
basin. 
  However, it is rather sensitive to other effects, as, for instance, 
 the space inhomogeneity introduced
by the grid used to integrate the Hartree and exchange-correlation 
energies in our ab initio method~\cite{Jose}.
  On the other hand, the preconditioned  CG method appears to be
more robust against this kind of effects.

\begin{figure}
\includegraphics[width=\columnwidth]{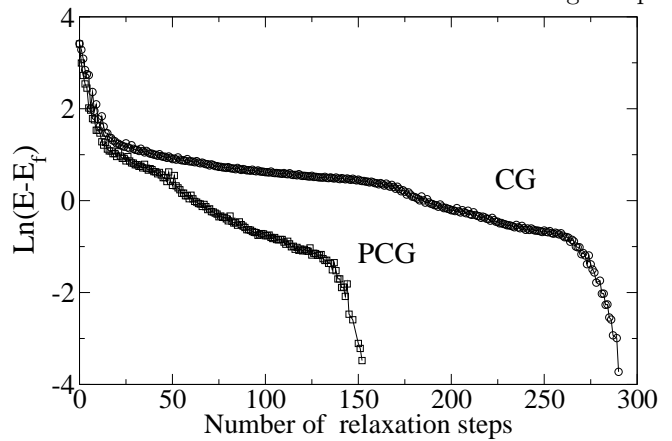}
\caption[3]{\label{bio} 
   Logarithmic convergence of the energy of a two-base-pair DNA strand 
relaxed with conventional conjugate gradients (CG) and preconditioned
conjugate gradients (PGC) methods, using atomic forces obtained from
density functional theory \cite{Jose}. 
}
\end{figure}

  Fig.~\ref{bio} shows the convergence for a piece of a double
helix of DNA with two base pairs (134 atoms).
  The forces in this case were calculated using ab initio 
density-functional theory, norm-conserving pseudopotentials 
and a basis set of numerical atomic orbitals.\cite{Jose}
  The preconditioned conjugated-gradient method represents a much 
better option for relaxing this molecular system. 
  Low energy modes not included in the acoustic branch are responsible
of the ill-conditioning on the relaxation of these sort of structures 
and the model takes account of them, improving the convergence by
a factor of two.


  We have developed two methods for accelerating first principles
structure relaxations, based on a classical and universal model
Hessian.
  Drastic improvements in overall efficiency are achieved, reducing the 
number of minimization steps by factors of 2 to 8 in the cases studied.
  Of the two methods presented, the Broyden method is more efficient
when sufficiently close to the minimum, while the preconditioned
CG method is more robust when the energy landscape is far from
harmonic.

\acknowledgments
  This work was funded by the Spanish Ministerio de Ciencia y
Tecnolog\'{\i}a under grant BMF2000-1312 and by the Fundaci\'on
Ram\'on Areces.
  MFS acknowledges support from a Studentship from the Comunidad
Aut\'onoma de Madrid.

\end{document}